# Where are GIScience Faculty Hired from? Analyzing Faculty Mobility and Research Themes Through Hiring Networks

Yanbing Chen [1,2], Jonathan Nelson [1], Bing Zhou [2,3], Ryan Zhenqi Zhou [2,4], Shan Ye [2,5], Haokun Liu [2,6], Zhining Gu [2,7], Armita Kar [2,8], Hoeyun Kwon [2,9], Pengyu Chen [2,10], Maoran Sun [2,11], Yuhao Kang [2,12]

[1] Department of Geography, University of Wisconsin-Madison

[2] GISphere Corporation

[3] Department of Geography, Pennsylvania State University

[4] Department of Geography, University at Buffalo, The State University of New York

[5] School of Information Engineering, China University of Geosciences (Beijing)

[6] Institute of Geography and Sustainability, University of Lausanne

[7] School of Geographical Sciences and Urban Planning, Arizona State University

[8] Department of Geography and Geoinformation Science, George Mason University

[9] Department of Earth, Environmental, and Geospatial Sciences, Lehman College, The City University of New York

[10] Department of Geography, University of South Carolina

[11] Department of Architecture, University of Cambridge

[12] GISense Lab, Department of Geography and the Environment, The University of Texas at Austin

[*] Corresponding author: Yuhao Kang, yuhao.kang@austin.utexas.edu

**Abstract**: Academia is profoundly influenced by faculty hiring networks, which serve as critical conduits for knowledge dissemination and the formation of collaborative research initiatives. While extensive research in various disciplines has revealed the institutional hierarchies inherent in these networks, their impacts within GIScience remain underexplored. To fill this gap, this study analyzes the placement patterns of 946 GIScience faculty worldwide by mapping the connections between PhD-granting institutions and current faculty affiliations. Our dataset, which is compiled from volunteer-contributed information, is the most comprehensive collection available in this field. While there may be some limitations in its representativeness, its scope and depth provide a unique and valuable perspective on the global placement patterns of GIScience faculty. Our analysis reveals several influential programs in placing GIScience faculty, with hiring concentrated in the western countries. We examined the diversity index to assess the representation of regions and institutions within the global GIScience faculty network. We observe significant internal retention at both the continental and country levels, and a high level of non-self-hired ratio at the institutional level. Over time, research themes have also evolved, with growing research clusters emphasis on spatial data analytics, cartography and geovisualization, geocomputation, and environmental sciences, etc. These results illuminate the influence of hiring practices on global knowledge dissemination and contribute to promoting academic equity within GIScience and Geography.

# 1. Introduction

Academia is profoundly shaped by the dynamics of faculty hiring networks, acting as a pathway for knowledge dissemination and collaborative research formation (Curran et al., 2020). Studies in various disciplines have highlighted the complexities of faculty hiring networks, with particular attention to how prestige and institutional hierarchies shape academic careers (Clauset et al., 2015; C. A. Lee, 2022). Prestige, in this context, typically refers to whether a faculty member earned their degree from a high-ranked institution, which can significantly influence job placement. Clauset et al. (2015) analyzed the career trajectories of 19,000 faculty members across business, computer science, and history using network science, revealing a *hierarchical nature* in faculty hiring. This hierarchy refers to a system where top-ranked institutions tend to hire from similarly prestigious institutions, creating a closed loop that limits opportunities for graduates from lower-ranked institutions. Wapman et al. (2022) extended this analysis to a larger dataset of 295,089 U.S. faculty members across 107 fields, confirming that these hierarchical patterns persist across disciplines. Their findings revealed widespread inequality in faculty output, with 80% of domestically educated faculty coming from just 20.4% of universities. Such trends perpetuate a concentration of academic power within a small subset of institutions, limiting the diversity of thought and research that can emerge from less prestigious institutions. Geographic Information Science (GIScience), as a discipline, has experienced significant growth in recent decades, with a marked increase in the recruitment of specialized GIS faculty. However, studies on faculty hiring patterns within the field remain limited, and it is unclear whether the hierarchical hiring structures observed in other disciplines also persist while hiring in GIScience.

The patterns of faculty hiring can significantly affect which research themes dominate specific disciplines and how new ideas are circulated. Terviö (2011) indicates that the division within economics is especially pronounced, reflecting the well-documented "freshwater" (Emphasizes market efficiency, minimal government intervention; linked to Chicago school of economics) and "saltwater" schools of thought (Advocates for government intervention, Keynesian economics; linked to coastal universities such as Harvard University), and has remained stable over time. Such divisions continue to perpetuate inequities in knowledge transmission, as PhD graduates carry ideas from their original institutions to new ones, thereby influencing the academic discourse and research directions in their new environments (C. A. Lee, 2022). Wapman et al. (2022) explored the hiring networks in United States and found that institutional prestige heavily influences faculty placement, thereby perpetuating a cycle where prestigious departments continue to dominate the field's research output and thematic focus. This movement of ideas, while potentially enriching for the receiving institution, also reinforces existing academic hierarchies, limiting the diversity of perspectives within the field. Currently, few studies focus on the dynamics of knowledge exchange through hierarchical faculty hiring networks.

Hiring networks not only determine who gets hired but also shape research agendas and academic practices by determining who holds gatekeeping roles, such as editors and contributors in influential journals. As Schurr et al. (2020) point out, the underrepresentation of women in these positions in Geography affects the thematic direction of research, often privileging traditional topics over more diverse or intimate perspectives. Franklin et al. (2021) further highlight the persistent underrepresentation of women in editorial teams of quantitative geography journals. Additionally, the lack of international representation in gatekeeping roles reinforces a narrow, Anglophone-dominated perspective, limiting the diversity of voices from non-core countries and impacting the global reach of geographical research. The concentration of power in hiring networks perpetuates inequalities, affecting the distribution of opportunities within the field.

Research on academic hiring networks has also traditionally been constrained by regional focus, with significant studies emerging from countries and regions such as Europe (Musselin, 2004), Japan (Horta et al., 2011), the United States (Clauset et al., 2015), Canada (Nevin, 2019), and Italy (Gallina et al., 2023). These investigations, while providing valuable insights into the dynamics of academic hiring within specific national boundaries, often do not capture the complexity of global academic mobility and collaboration. The emphasis on national contexts has left a gap in understanding how academic networks operate on a worldwide scale, particularly in fields that are inherently global in nature.

While factors influencing faculty hiring and the resulting academic environment have been extensively investigated in select fields like economics (Orland & Padubrin, 2022), communication (Mai et al., 2015), computer science (Way et al., 2016), urban planning (C. A. Lee, 2022), mathematics (FitzGerald et al., 2023), and geography (Liu et al., 2024), such insights are notably lacking in the context of GIScience. As an emerging, globally connected field within Geography, GIScience has undergone significant expansion over the past three decades (Buttenfield, 2022; Egenhofer et al., 2016). Moreover, GIScience is highly international, as the growing number of GIS programs has attracted a significant number of international students (Y. Wang et al., 2023). As a result, there has been a major increase in the recruitment of GIS faculty, specializing in cartography, spatial analysis, remote sensing, spatial data science, and GeoAI, among other topics; yet, very little is known about GIScience faculty hiring, placement, and networks. To address this gap in understanding, this article explores the global mobility patterns, research themes, and academic dynamics of GIScience faculty, and aims to answer the following two interrelated research questions:

1. What are the global placement patterns of GIScience faculty?
2. How are the research themes among GIScience faculty distributed, and how do they evolve over time?

Our study aims to extend the analysis of hiring network outcomes beyond the confines of national academic hiring networks and well-documented dimensions of gender and affiliation with elite institutions. We examine how these networks facilitate or hinder the spread of ideas, impacting innovation within GIScience. Specifically, we leverage the GISphere Project[1], an extensive database that collects, maintains, and shares up-to-date information on global GIS graduate programs and faculty. This data source allows us to examine faculty hiring in GIScience not just at the national level but also from a global perspective, providing information on individual research interests and departmental strengths. It is important to note that our dataset relies on volunteer-contributed information, which inherently constrains its representativeness—particularly for non-Western countries, where data are comparatively underrepresented. Nonetheless, given the substantial volume of data collected, our study effectively captures the picture of the academic landscape in GIScience, and could advance our knowledge of diverse geographic and institutional contexts within the field. By aggregating detailed information on faculty's educational backgrounds and work experiences from diverse sources, including ORCID, LinkedIn, personal websites, and other online platforms, we create a global GIS faculty network graph, visually representing the connections between PhD-granting institutions and current academic affiliations. Building on the observed global placement pattern, we employ natural language processing (NLP) to assess the significance of the hiring network in facilitating the transmission of knowledge.

The article is organized as follows: Section 2 situates the expansion of GIS education within a broader literature focused on academic hiring patterns in other fields. Section 3 describes the data collection and preprocessing. Section 4 presents the results to demonstrate global GIS faculty placement patterns. Section 5 discusses the use of NLP and word cloud visualization in detecting research clusters based on research themes and the hiring network within GIScience.

---

[1] https://gisphere.info/

Section 6 offers insights and implications for GIS scholars and educational institutions, providing a comprehensive understanding of our findings' scope and contextualizing our research within the broader GIScience and geography framework.

## 2. Literature Review

This section explores the existing body of research on faculty hiring patterns across academia and examines the growth of GIScience as an academic field. By analyzing these two topics together, we aim to understand how the broader academic hiring trends influence and are reflected in the specific context of GIScience.

### *2.1 Faculty Hiring Studies Across Academia*

Academic hiring trends have been a subject of extensive study across various disciplines. In recent years, there has been growing interest in quantifying the faculty hiring process to examine inequity in academia (Dewidar et al., 2022; Fernandes et al., 2020; E. Lee et al., 2021; Mai et al., 2015; Orland & Padubrin, 2022; Way et al., 2016; Wu et al., 2023; Xierali et al., 2021). Clauset et al. (2015) pointed out that the faculty hiring process in academia is heavily influenced by the prestige of academic institutions. They discovered a "closed doctoral ecosystem" where prestigious institutions predominantly hire graduates from similarly elite universities, perpetuating a cycle that limits diversity of ideas and innovation in academic fields. Wapman et al. (2022) further highlighted this trend, revealing that five U.S. universities have trained 1-in-8 tenure-track faculty members in the US. The results are consistent with findings from a range of disciplines. According to Burris (2004), 32% of sociology faculty in the U.S. graduated from the top five sociology programs. Similarly, 34% of political science faculty and 39% of history faculty come from their respective top five programs. In computer science, only 12 percent of faculty were able to secure jobs at universities more prestigious than where they graduated, a number that drops to 6 percent in economics (Wapman et al., 2022). In Geography (Liu et al., 2024), there is a prevalent pattern of downward mobility within a hierarchical academic structure, where scientists often move to institutions of equal or lower prestige. In the meanwhile, the spatial mobility of geography scientists is dominated by North America, Western and Northern Europe, East Asia, and Oceania, with increasing inequality and multipolarity over time.

Beyond institutional type and ranking, additional factors influencing academic career trajectories include gender (Weisshaar, 2017), race/ethnicity (Grier & Poole, 2020), advisor sponsorship (Pinheiro et al., 2017), and postdoctoral training (Lin & Chiu, 2016). It is worth noting that postdoctoral positions have emerged as a critical component of the post-PhD labor market, enabling scholars to expand their professional networks and often serving as a vital stepping stone toward securing tenure-track positions. Meanwhile, many scholars follow non-traditional career paths—such as roles in government, industry, visiting professorships, or non-tenure-track positions—before establishing a tenure-track position. Given the consistency of available data for scholars' educational background and career trajectories, our study focuses exclusively on PhD-granting institutions; a comprehensive examination of the other factors remains a promising avenue for future research.

*2.2 Expansion of GIScience as an Academic Field & Profession*

In recent years, the rapid growth in GIS, fueled by technological advances and the demand for spatial analysis, is revolutionizing data-driven solutions across various fields (Longley, 2000). Notably, GIS enhances agriculture through crop planning and soil analysis (Zhang & Cao, 2019), aids in managing healthcare resources and public health planning in epidemiology (F. Wang, 2020), streamlines routing and traffic in transportation (Bruniecki et al., 2016), and improves disaster modeling and response (Abdalla & Tao, 2005; Tomaszewski, 2020). According to the Bureau of Labor Statistics (*Surveying and Mapping Technicians*, 2024), demand for GIS professionals is projected to rise in diverse sectors, with a six percent job growth anticipated between 2023 and 2033. Moreover, the North American GIS market is expected to expand by USD 11.4 billion, growing at a CAGR (Compound Annual Growth Rate) of 23.7% from 2024 to 2029, as reported by Technavio (2025). This evolution has prompted educational institutions to not only increase the number of GIS programs but also to enhance them by incorporating more technique-oriented courses to meet the growing demand for highly skilled professionals in the field. Thus, industry trends are reflected in academia, with a noticeable increase in GIScience research and publications. Since Michael F. Goodchild coined "geographical information science" (GIScience) in 1991, GIS publications quadrupled by 2020 (Wu et al., 2023), and the share of international collaborations in GIScience has more than tripled since 2000 (Biljecki, 2016). Additionally, the research scope has broadened into sub-domains like volunteer geographic information, geospatial data science, geovisualization, and so on. This expansion fosters an ongoing cycle that boosts funding and the recruitment of more GIS-focused scholars in academia. Between 2021 and 2023, the National Science Foundation has awarded around 180 grants, amounting to over $83 million, for research related to GIS (*Mapping Science*, 2023). As a result of the flourish in the field, there has been a notable increase in the recruitment of specialized GIS faculty in the recent years. We noticed several institutions have initiated focused cluster hires in GIS faculties, such as the University of California Santa Barbara, University of Wisconsin-Madison, Texas A&M University, Peking University, University of Glasgow, and University of Tokyo.

Despite extensive research on hiring networks across various fields, GIScience, a crucial geography sub-domain for the past three decades, still lacks substantial studies in this area. This gap underscores the need for in-depth research into the academic mobility within GIScience. Such studies are crucial to determine if the faculty hiring process in GIScience also relies on the institutional hierarchy, affecting research diversity and innovation. Moreover, while research in other disciplines often focuses on hiring networks based on institutional data, many studies are limited to U.S. databases (FitzGerald et al., 2023; C. A. Lee, 2022; Wapman et al., 2022), excluding global data due to methodological challenges in obtaining it. This study created a comprehensive global database to analyze faculty hiring networks, advancing understanding of academic mobility in GIScience.

## 3. Data Collection and Preprocess

### 3.1 Data Source

The dataset employed in this study is sourced from the GISphere Guide[2], which provides comprehensive information on approximately 2,065 faculty specializing in Geographic Information Science (GIS) and related fields worldwide (as of July 2023). The repository covers 480 geography graduate programs (offering master's and doctoral degrees) across 96 countries and regions, with a substantial proportion of faculty categorized into one or more of seven sub-disciplines: Geographic Information Science (GIS), Satellite Navigation (GNSS), Remote Sensing (RS), Physical Geography, Human Geography, Urban Planning, and Transportation. The dataset includes current institutional affiliations, appointment details, research interests, and specific areas of expertise (Y. Wang et al., 2023). Data were collected by dedicated GISphere project volunteers between 2019 and 2024 via a combination of crowdsourcing and active searching, and the information is subject to annual verification and updates to ensure accuracy.

### 3.2 Data Preprocessing

Our focus is narrowed to faculty who specialize in the GIS domain. To accurately identify GIS faculty within the broader academic community, our data cleaning and inclusion process proceeded as follows

---

[2] https://gisphere.info/

(

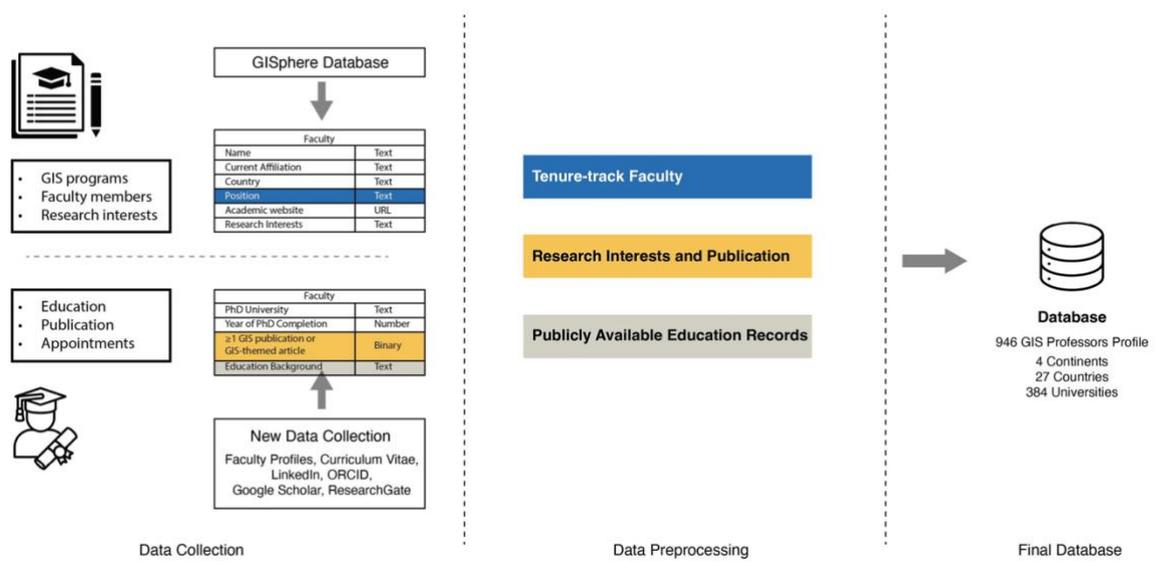

Figure 1):

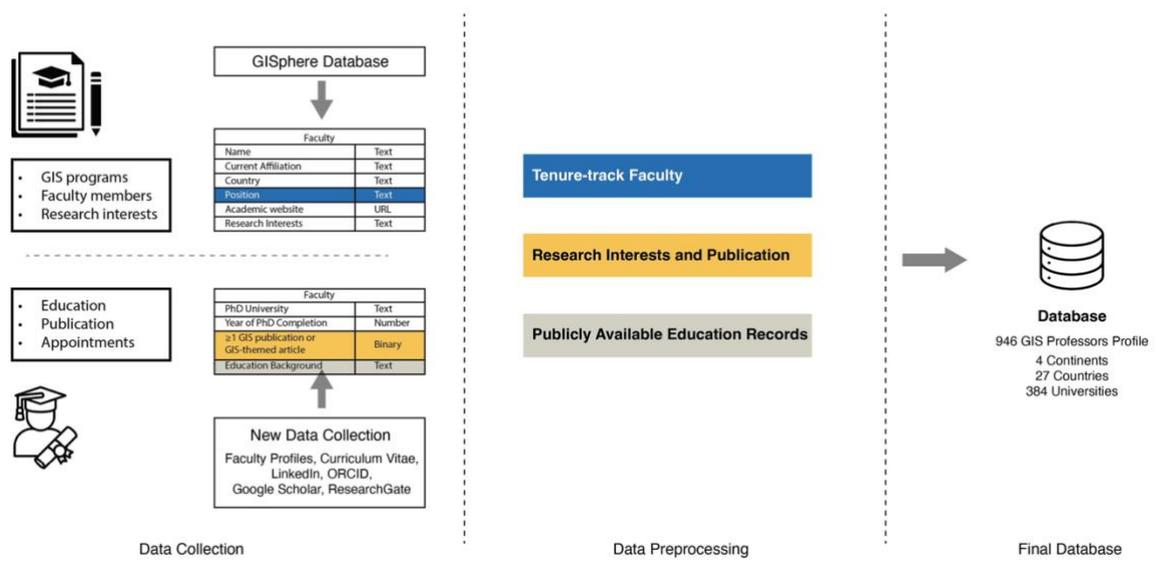

*Figure 1. Workflow of data collection and preprocessing.*

**Tenure-Track Faculty.** GISphere Guide classifies potentially GIS-related faculty into seven sub-disciplines, a categorization that reflects the interdisciplinary essence of GIScience. For all faculty listed in the GISphere Guide, our analysis was restricted to tenure-track faculty — a common practice in studies of faculty hiring (Clauset et al., 2015; C. A. Lee, 2022; Wapman et al., 2022). In accordance with the American Association of University Professors' definition of tenure-track positions (*Tenure and Teaching-Intensive Appointments*, 2010) and based on publicly available information from faculty profiles, we excluded individuals whose titles explicitly indicated that they held non-tenure-track roles or retired, such as Emeritus faculty, professors of practice, adjunct professors, professors of teaching, and research professors. While this approach ensures consistency, we acknowledge that some ambiguity may persist where departmental practices or titles are not standardized or unclear. Moreover, this exclusion represents a limitation of the study, as it may overlook the significant contributions of these roles to academic research, teaching/mentoring, and service. However, our focus remains on tenure-track faculty because these positions are typically associated with long-term institutional investment, formalized expectations of research productivity, and a central role in shaping the academic pipeline through hiring, promotion, and curriculum development. While research and teaching professors may also mentor PhD students and serve as Principal Investigators, their institutional roles and career trajectories often differ from those of tenure-track faculty, particularly in terms of departmental governance, tenure security, and influence on faculty hiring and promotion processes. Thus, our decision to focus on tenure-track faculty ensures a consistent format and aligns with the study's aim of examining patterns in faculty hiring and academic career progression.

**Research Interests and Publication.** A key difficulty in filtering the GISphere dataset is distinguishing between faculty who focus on advancing GIScience and those who use Geographic Information Systems (GIS) as tools in various fields. To address this challenge, we implemented a two-step manual process. First, we examined research interests as listed in faculty profiles, curriculum vitae, and Google Scholar. A faculty member was included if their research interests explicitly contained GIS-related terminology (e.g., "geographic information systems," "spatial analysis," "geovisualization," "remote sensing"). These keywords were adopted based on the GIS research themes examined in Wu et al. (2022), who systematically parsed GIS journal content to compile a comprehensive list of relevant terminology. Second, we reviewed publication records available on Google Scholar, ResearchGate, and ORCID. Faculty were identified as GIScience scholars (i.e., those focused on advancing GIScience) if they had at least one publication in a recognized GIS-specific journal, or if the title and abstract of their publications in non-GIS-specific journals indicated a clear focus on GIS-related topics, according to the GIS journals and sub-domain topics defined by Wu et al. (2022). Faculty whose profiles and publication records did not meet these criteria were excluded from our analysis, indicating that they primarily use GIS as a tool.

**Publicly Available Educational Records.** Following the initial identification, we verified each professor's educational background and current affiliation through multiple sources (official departmental websites, LinkedIn profiles, personal webpages, curriculum vitae, ORCID profiles). In cases where faculty held appointments at multiple institutions, only the current, primary (tenure-track) affiliation was recorded. Furthermore, all tenure-track—including assistant, associate, and full professors (or their UK equivalents, such as lecturers and senior lecturers)—were grouped together to focus on overall placement patterns rather than rank-specific trends.

*3.3 Results*

By applying the criteria described above, we identified a final dataset comprising 946 GIS faculty members currently affiliated with 384 universities across 27 countries in North America, Europe, Asia, and Oceania. The manual data collection process was completed in March 2024, and subsequent updates were not included.

## 4. Placement Patterns in GIScience Faculty Hiring

*4.1 Network Analysis of Global GIS Faculty Placement*

In this section, we utilized network analysis to explore the influential GIS programs as defined by those that have the greatest impact on placing PhD graduates in faculty positions. Network analysis was employed to assess the global distribution of GIS graduate programs and identify key programs that produce and place graduates as faculties within the field. This method involves mapping and analyzing relationships between entities within a network, typically represented as nodes and edges (Hevey, 2018). Here, universities are depicted as nodes, and the links between these nodes represent the placement of graduates into academic positions.

Results from our network analysis revealed that the United States leads with 270 GIS faculty positions worldwide, comprising 28.54% of the total. China follows with 253 faculty members, accounting for 26.74%. The United Kingdom (N=79) and Canada (N=71) represent 8.35% and 7.51%, respectively. Germany (N=24), Japan (N=23), Switzerland (N=23), and Netherlands (N=23) each contribute over 2% to the total. The top eight countries with the most GIS faculty account for 80.97% (766 out of 946) of current GIS faculty positions.

Our global GIS faculty placement pattern identifies several influential institutions contributing to GIS programs (Figure 2). Wuhan University and Chinese Academy of Science, both in China, lead the way, each accounting for 5.50 percent and 3.38 percent of all GIS faculty, followed by University of California Santa Barbara (2.43%), The Ohio State University (2.33%), and Peking University (1.80%). Collectively, these five universities account for 15.43% of placements in GIS programs. Conversely, 50 universities placed only two faculty members, and 135 universities placed only one faculty member. Unlike other academic fields with more clearer hierarchies, GIS faculty placements are more evenly distributed, which indicates a relatively decentralized hiring pattern in GIS programs. However, it is worth noting that, for the top 25% (96 out of 384) of institutions, they have achieved a placement rate of 70.4%. Figure 3 additionally illustrates that the United States has a significant presence among the top 10 universities for faculty placement. The United States institutions also make up 40.63% (39 out of 96) of the top 25% of educational institutions in terms of GIS faculty placement. Figure 3, 4, 5 and 6 illustrates the faculty hiring networks for each continent by displaying the current affiliations of GIS faculty alongside the origins of their PhD degrees. To better illustrate the patterns of faculty mobility across institutions, we also created an interactive hierarchical edge bundling web visualization (link), which dynamically represents the academic hiring flows in GIScience.

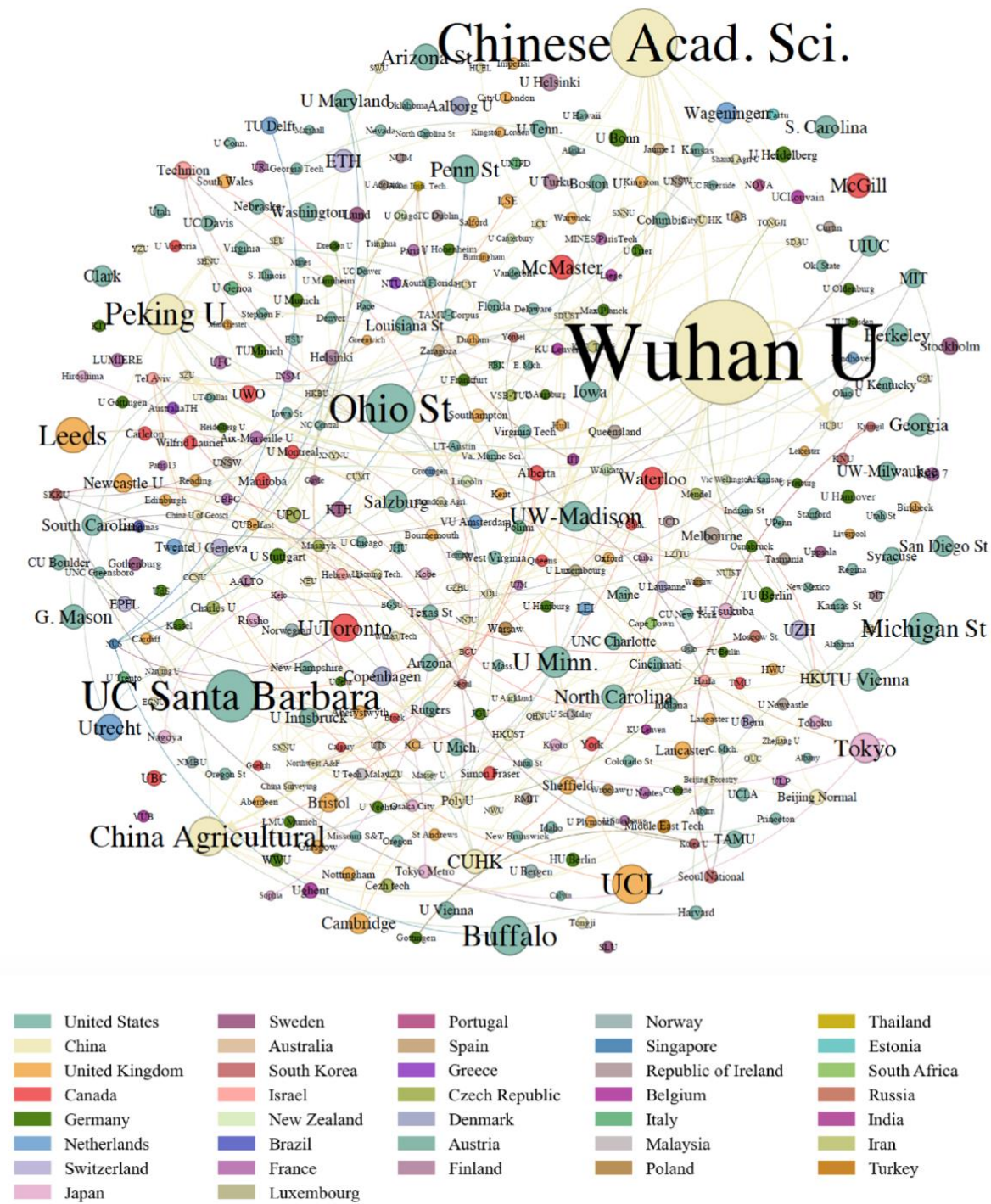

*Figure 2. Network graph of global GIS faculty placement. Each node represents a university,*

*and the colors of the nodes indicate the country of origin. The size of each node corresponds to the number of GIS faculty members originated from that university, with larger nodes indicating a higher number of faculty placements. The edges connecting the nodes represent the flow of GIS faculty between universities, with thicker edges indicating a greater number of faculty moving from one university to another.*

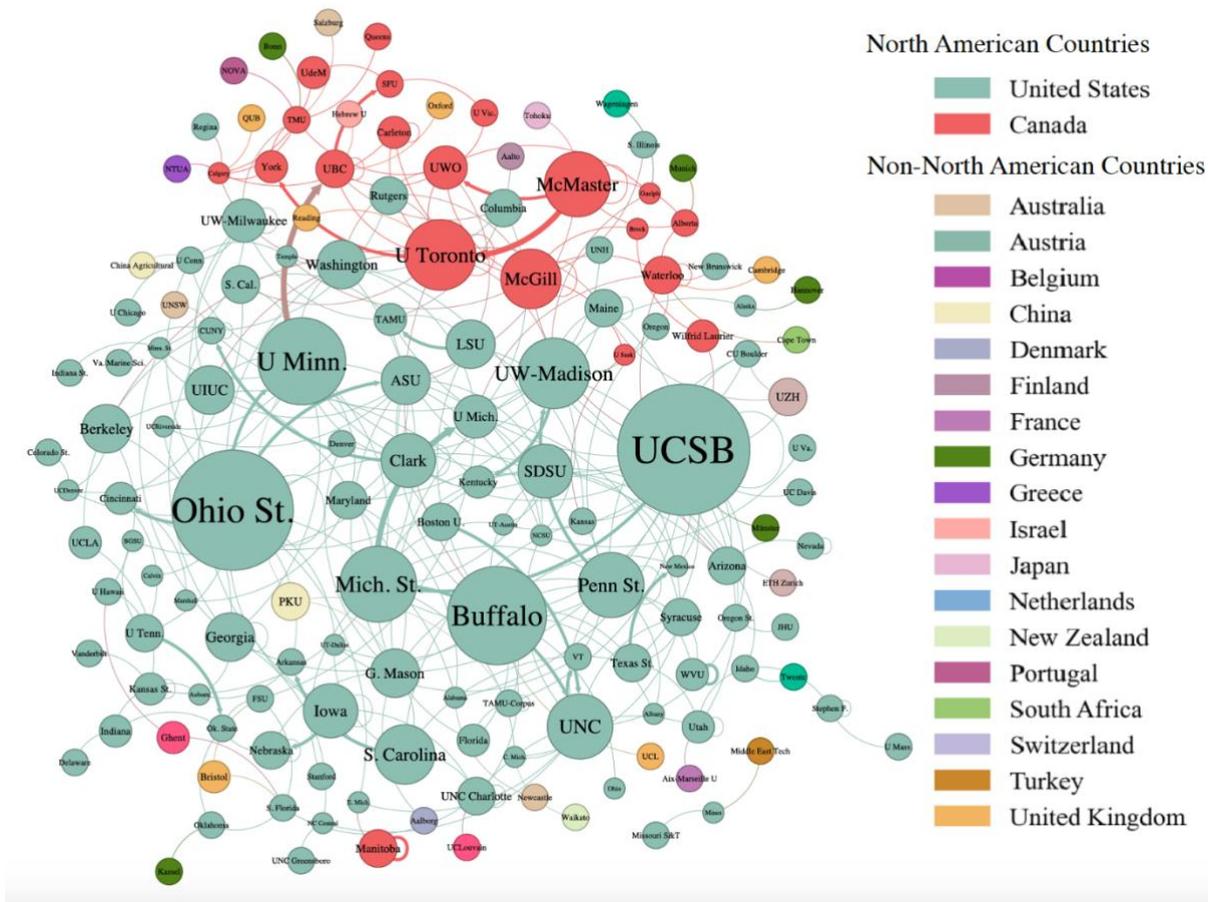

4(a). North American: Current GIS Faculty Affiliations and Their PhD Origins

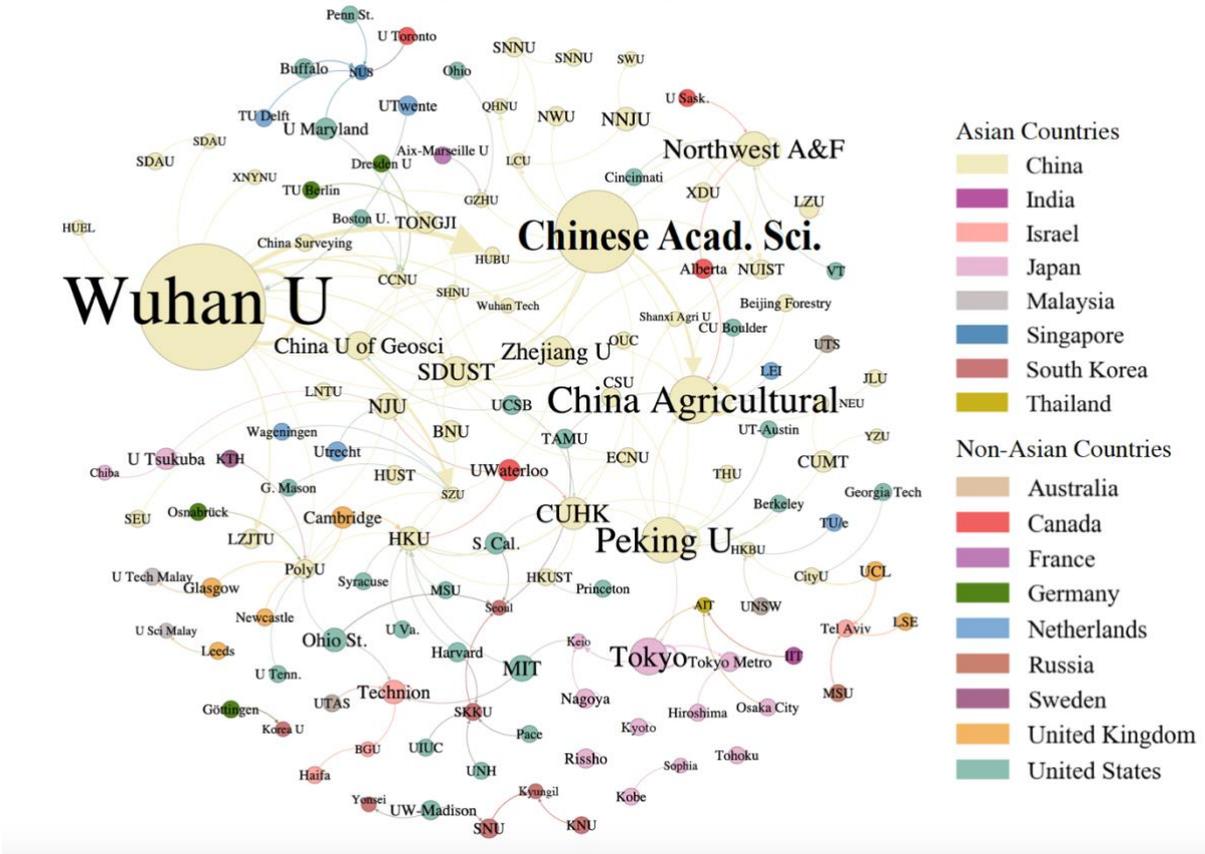

4(b). Asia: Current GIS Faculty Affiliations and Their PhD Origins

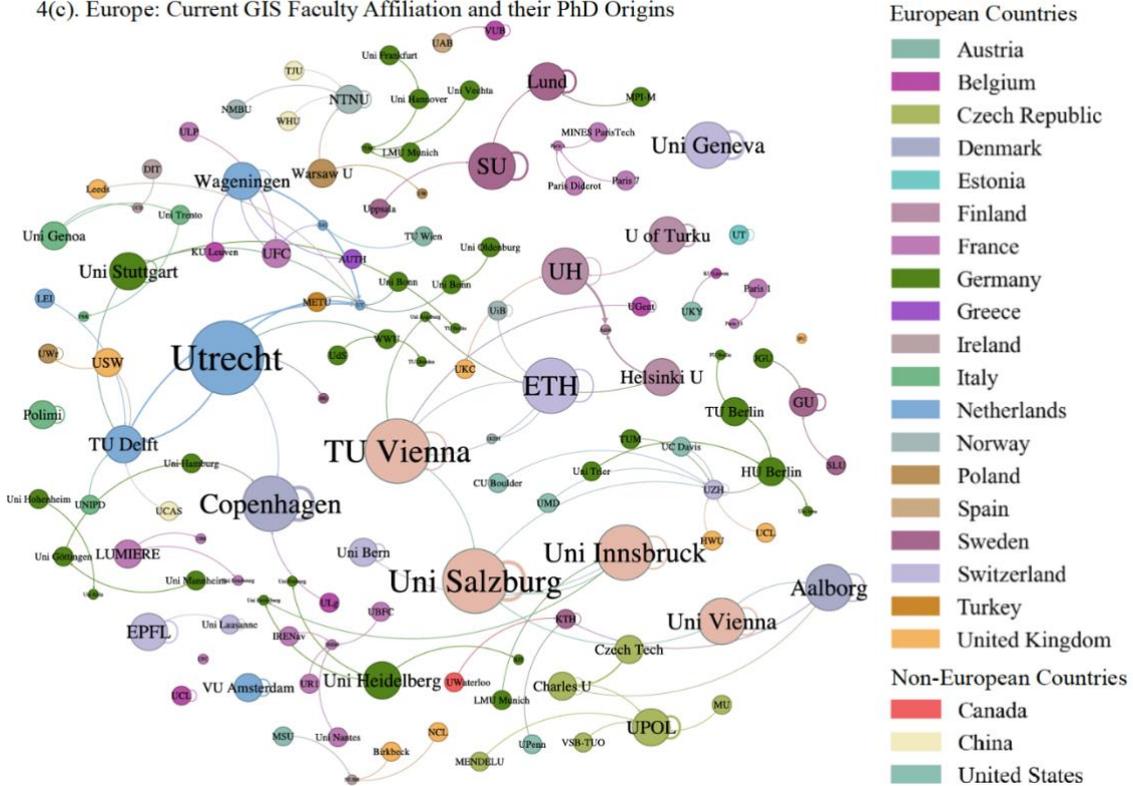

4(c). Europe: Current GIS Faculty Affiliation and their PhD Origins

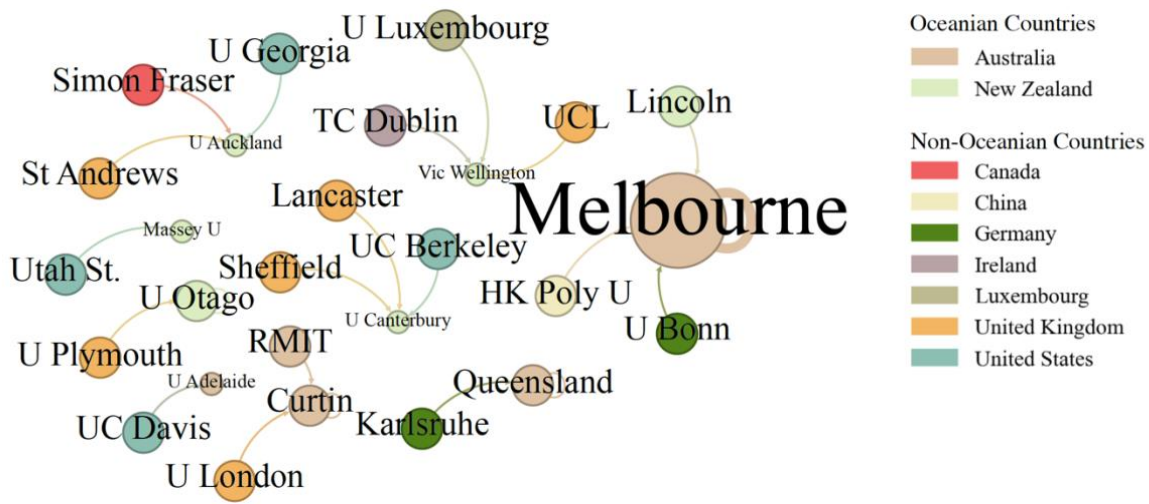

*Figure* 3. *Faculty Hiring Networks in North America. Figure 3 shows faculty currently employed in North American institutions and indicates whether their doctoral training was obtained in North American countries or Non-North American Countries.*

*Figure 4. Faculty Hiring Networks in Asia.*

*Figure 5. Faculty Hiring Networks in Europe.*

*Figure 6. Faculty Hiring Networks in Oceania.*

### *4.2 Diversity Index of Global GIS Faculty Placement*

This study introduces a new measure of diversity: the diversity index, based on the proportion of faculty originating from all other regions relative to the total affiliated faculty within a certain region. The calculation of the diversity index is not fixed to a specific numerator and denominator, acknowledging that different spatial scales can yield significantly divergent outcomes in geographical studies. Consequently, this investigation explores the diversity of academic networks at multiple spatial scales, including the continental, country, and institutional levels.

$$Diversity\ Index = \frac{Number\ of\ faculty\ originated\ from\ all\ other\ regions}{Number\ of\ faculty\ affiliated\ with\ certain\ region} \quad (1)$$

In this section, we describe the faculty placement patterns at three different levels —
*continental, country, and institutional* – as identified from our network analysis. The
continental level analysis reveals cross-border patterns in faculty mobility and hiring. The
country level analysis investigates the national distribution and diversity of GIScience faculty.
The institutional level analysis identifies more nuanced preferences in faculty hiring trends in
GIS disciplines. Our multi-level analysis aims to demonstrate the internal recruitment ratio at
different scales and how geographical location affects academic opportunities in the field, and
all the figures in Section 4.2 can be found in the interactive dashboard of GIS faculty hiring [3].

*4.2.1  Diversity Index at Four Continents*

Among the 946 GIS faculty reflected in the dataset, 345 faculty members are currently affiliated with institutions in North America, 316 in Asia, 260 in Europe, and 25 in Oceania. Of all the faculty, 40.49% completed their PhD degrees in North America, 30.87% in Europe, and 26.74% in Asia (Figure 4). The remaining 18 faculty members include 16 who completed their degrees in Oceania, one in South America, and one in Africa.

---

[3] https://public.tableau.com/app/profile/6068/viz/GISFacultyHiringNetwork/continent-merimekkochart

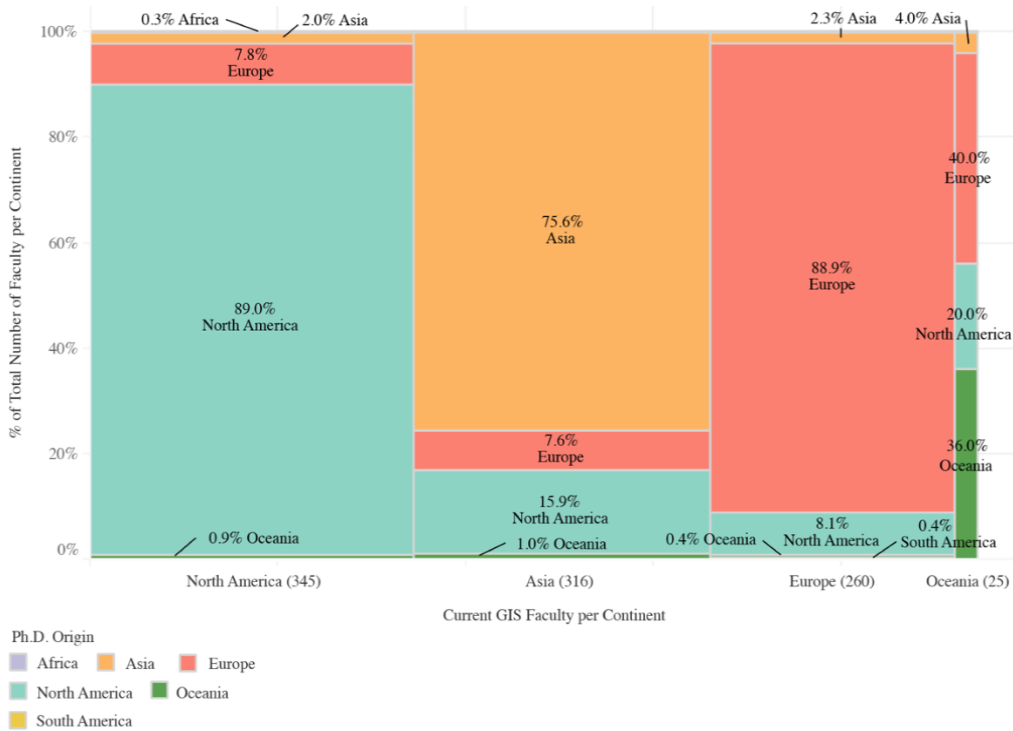

*Figure 4. Marimekko chart visualizing the distribution of GIS faculty placements across four continents (North America, Asia, Europe, and Oceania) and their PhD origins. The X-axis represents the total number of current GIS faculty in each continent, while the Y-axis shows the percentage of faculty based on where they obtained their Ph.Ds. The width of each segment on the X-axis corresponds to the number of GIS faculty in that continent, and the colored segments represent the proportion of faculty whose PhD origins are from different continents. For example, a total of 345 GIS faculty members are currently affiliated with institutions in North America. Of these, 88.99% obtained their PhDs in North America, 7.83% in Europe, 2.03% in Asia, 0.87% in Oceania, and 0.29% in Africa.*

The diversity index at each continent is calculated by dividing the number of faculty originated in all other continents by the number of faculty affiliated with that continent. The diversity index remains relatively low, at 24.37% for Asia, 11.15% for Europe, and 11.01% for North America, with all values falling below 25%. Oceania, smaller in total number of current GIS faculty, prefers recruitment from European institutions (10) and intercontinental placement (9).

These two groups together account for 76% of the faculty origins in Oceania. This indicates a strong preference for internal retention within the continental-level hiring networks.

We further analyzed the geographic origins of PhD graduates by continent, showing placement patterns as displayed in Figure 5. In North America, GIS faculty predominantly hold Ph.Ds. from the United States (257) and Canada (50), complemented by graduates from the U.K. and China. In Europe, the U.K. is the prominent source (64), with other European countries also contributing to the mix. In Asia, the hiring pattern is centralized in China (202), with Japan (24) as a secondary source, and a few representations from Western-educated faculty. For Oceania, Australia is the main provider of GIS PhDs (7), alongside a handful from the U.K (6) and the U.S. (4).

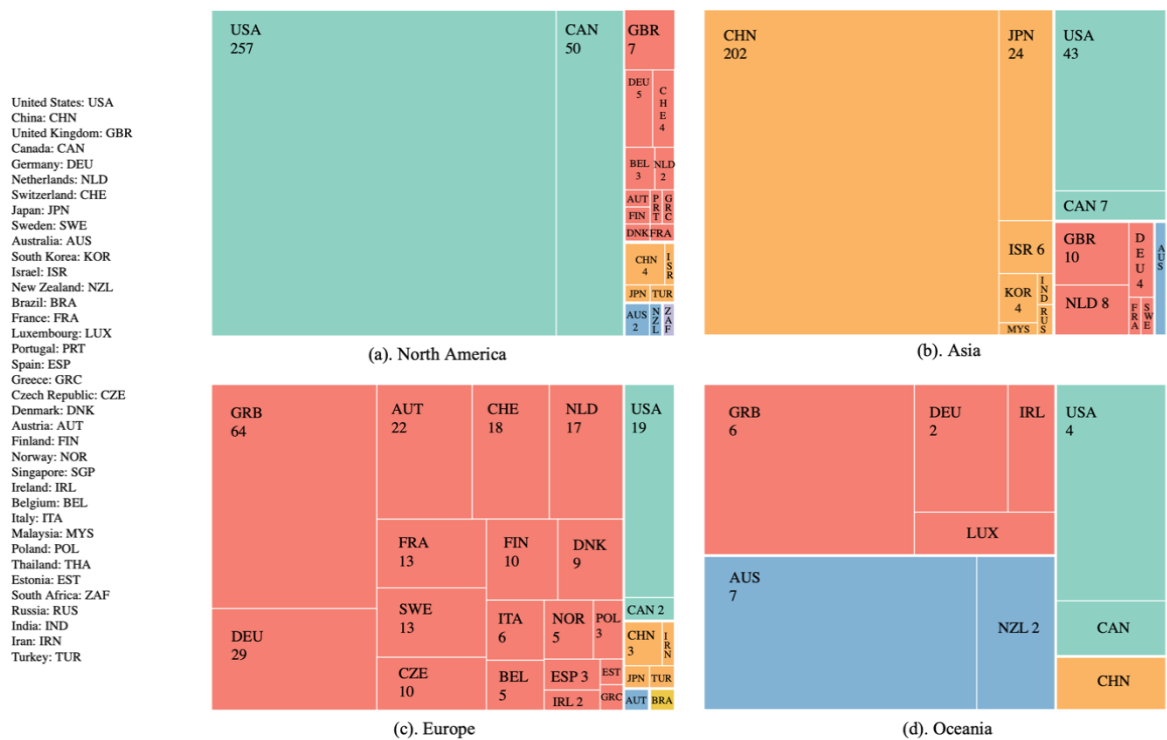

*Figure 5. Tree maps of faculty placement by country. These tree maps visualize the distribution of GIS faculty placements across various countries, segmented into four continents: (a) North America, (b) Asia, (c) Europe, and (d) Oceania. Each rectangle within the treemap represents a country, with the size of the rectangle proportional to the number of*

*GIS faculty placements in that continent. The colors distinguish different continents. The country code labels each rectangle, and the number inside each rectangle indicates how many faculty members who completed their PhD in that country are currently affiliated with institutions within that continent. Rectangles without numbers represent countries with only one faculty placement.*

*4.2.2   Continental-level Diversity Index*

At the continental level, the diversity index for each country is calculated by dividing the number of faculty whose PhDs were obtained in countries outside their continent by the total number of faculty affiliated with institutions in that country. The continental level diversity index highlights the extent to which countries rely on talent from other continents. A higher index at this level indicates significant international mobility. The average diversity index is 14.27%; in other words, countries, on average, hire only 14.27% of their faculties with a PhD earned from other continents (*Figure* 6). Only eight countries exceed this average: five from Asia, two from Oceania, and one from Europe. Except for China, all these countries have 15 or fewer GIS faculty members. In Table 1, except for Singapore (100%), New Zealand (90.91%), South Korea (73.33%), and Malaysia (66.67%), all other countries show a relatively low diversity index, indicating that their hiring decisions tend to favor domestic or regionally local candidates over international applicants.

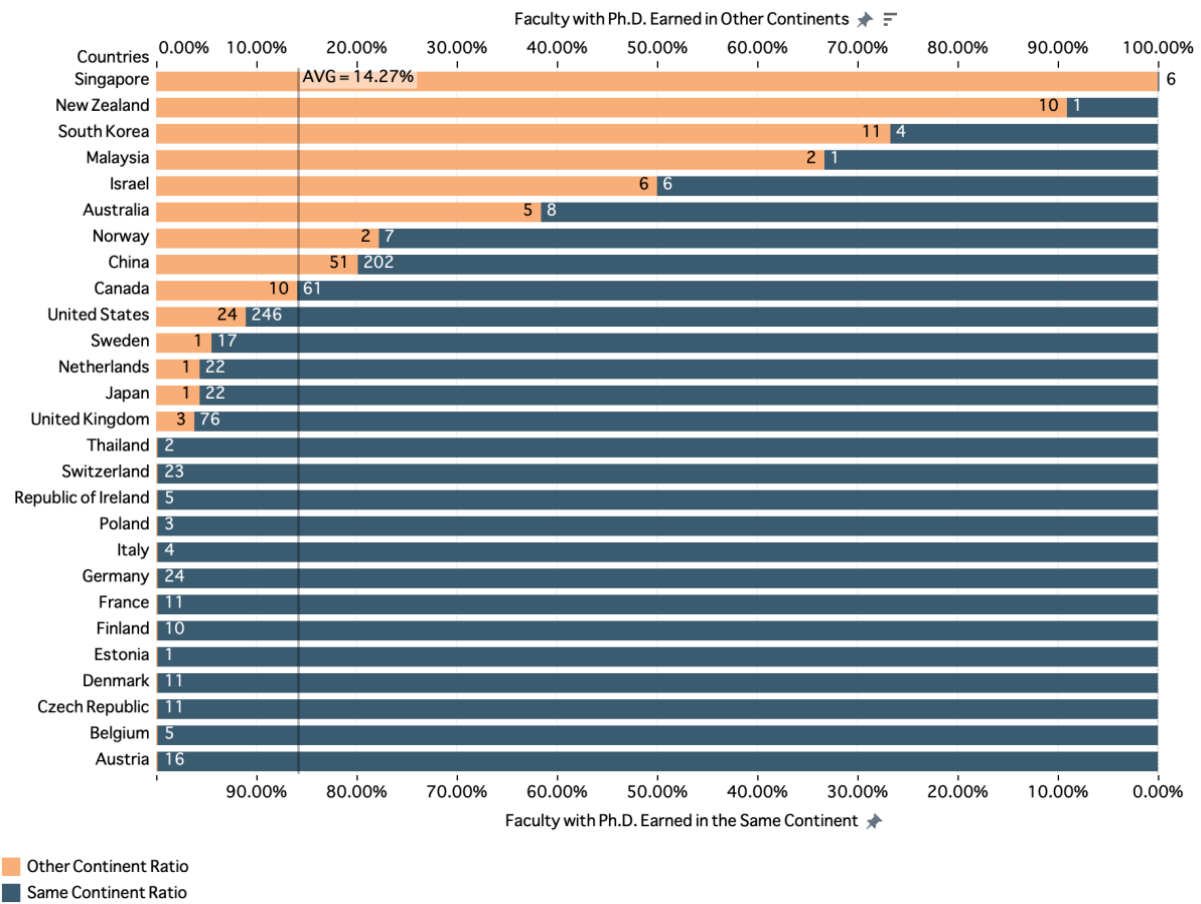

*Figure* 6. *Diversity index at the continental level. This figure shows the distribution of GIS faculty across 27 countries, divided by the location where they earned their PhD degrees. The average continental-level diversity index is 14.27%, with countries ranked by this ratio, starting with Singapore, down to Austria. The horizontal bars represent the percentage of faculty in each country who earned their PhD in either the same continent as their current institution (blue) or a different continent (orange). Black numbers represent faculty with PhDs from other continents, and white numbers show those with PhDs from the same continent.*

### 4.2.3   Country-level Diversity Index

At the country level, the diversity index for each country is calculated by dividing the number of faculty whose PhDs were awarded in countries other than the country itself by the total number of faculty affiliated with that country. Figure 7 demonstrates varying degrees of reliance on domestic versus international academic expertise in the field, with an average

Diversity Index of 25.21%. Countries like Singapore and Thailand stand out with the highest diversity, where all of their GIS faculty have obtained PhD internationally. Conversely, France, Italy, and Estonia show the lowest diversity, with their faculties comprised entirely of domestic PhD holders. Middle-ground countries such as the United States and China have more international representation, with diversity index ratios of 12.30% and 20.50%, respectively. The countries with the highest diversity indices tend to have a relatively low number of faculty positions; eight countries with a diversity index over 50% do not exceed 15 faculty members. Meanwhile, the top four countries with the most GIS faculty demonstrate modest or low international representation: the US at 12.30%, China at 20.55%, the UK at 29.21%, with only Canada showing a higher rate at 42.75%.

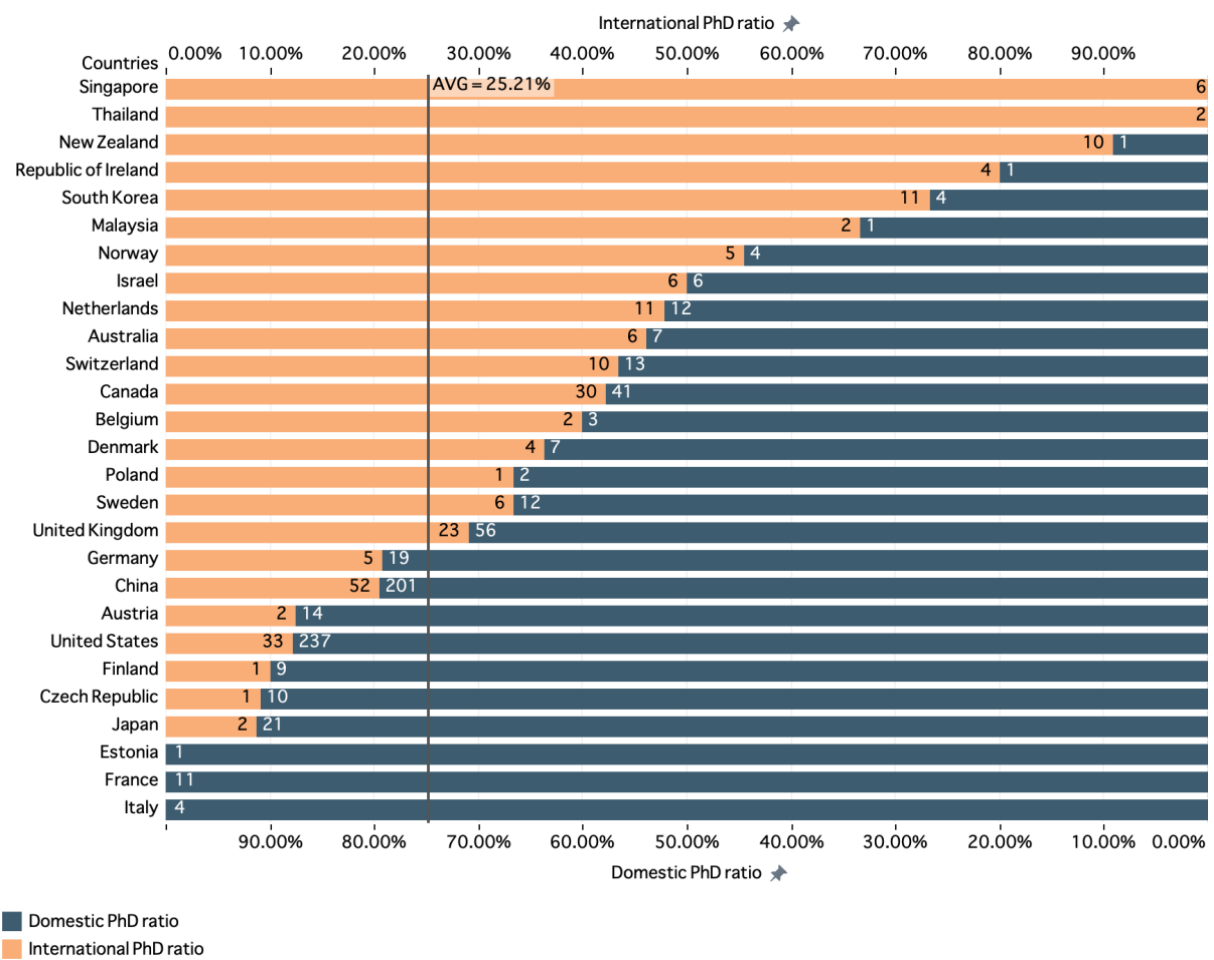

*Figure* 7. *Domestic PhD Ratio and International PhD Ratio for each Countries. This figure*

*shows the distribution of faculty by PhD origin across 27 countries, highlighting domestic versus international PhD ratios. The average international PhD ratio is 25.21%, with countries ranked by this ratio, starting with Singapore, which has the highest percentage of internationally educated faculty, down to Italy. Blue bars represent the percentage of faculty with PhDs earned domestically, while orange bars show those with international PhDs. Black numbers on orange bars indicate the count of international PhD holders, and white numbers on blue bars show domestic PhD holders.*

*4.2.4 Institutional Diversity Index*

At the institutional level, the diversity index for each institution is calculated by dividing the number of faculty whose PhDs were obtained from institutions other than the current one by the total number of faculty affiliated with that institution. We examined the diversity index for every institution and grouped by countries, indicating an institution's inclination to hire faculty with PhDs from outside its own institutions, which can be understood as the non-self-hired ratio (Wapman et al. 2022). The average diversity index is 60.66%. In Figure 8, Institutions in South Korea, Singapore, Republic of Ireland and Thailand show 100% non-self-hired ratio. Conversely, Estonia, Austria, Denmark, Japan and Belgium show high internal retention, exceeding 60% of their faculty having earned their PhDs from the same institutions where they are currently affiliated. The top four countries with the highest number of GIS faculty positions all exhibit a strong preference for non-self-hiring. In North America, the United States and Canada display external hiring ratios of 94.59% and 85.78%, respectively. The U.K. follows with an external hiring ratio of 77.22%, and China shows a relatively higher preference for internal retention among these countries, at 73.52%.

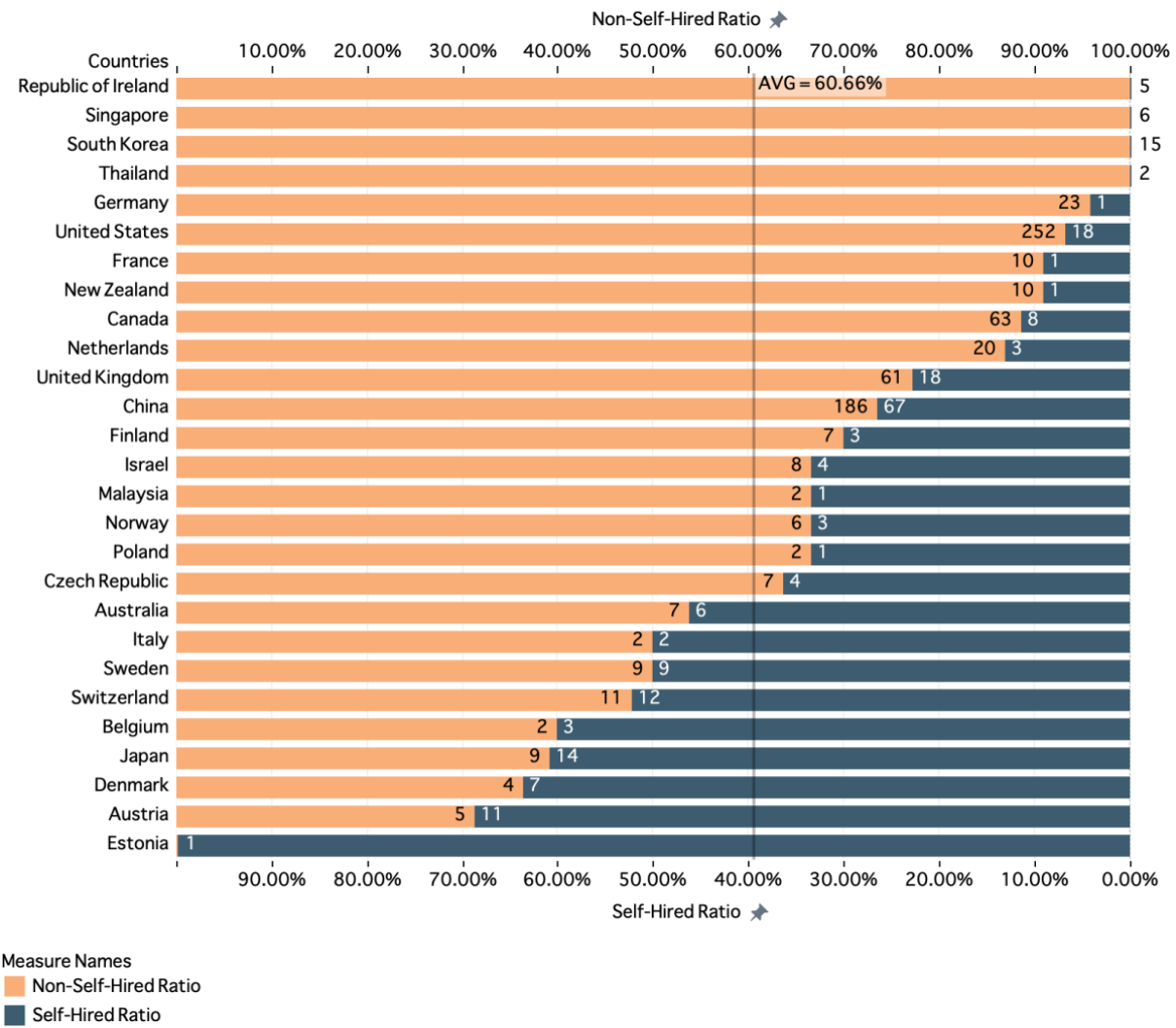

*Figure 8. Self-hired ratio and non-self-hired ratio in each country. Orange bars represent the non-self-hired ratio, while blue bars represent self-hired ratio. Black numbers on the bars show the count of externally retained faculty, white numbers represent internally retained faculty. The chart is ordered by non-self-hired ratio, with countries like the Republic of Ireland and Singapore having the highest non-self-hired ratio. The average non-self-hired ratio 60.66%.*

### 4.2.5   *Comparison of Diversity Index at Different Level*

The diversity index at different levels—continental, country, and institutional—provides insights into the geographic and institutional preferences in academic hiring at different regions (Table 1). It is worth noticing that the difference between the continental and country-level diversity indices reveals not only the degree of international diversity but also the influence of neighboring countries in faculty hiring. For example, 10 European countries exceed the average country-level diversity index (25.21%), but the number drops to 4 when comparing the average continental level diversity index (14.27%), and 9 European countries show 0% at the continental level. This indicates that most countries in Europe tend to hire PhD graduates from within the continent even if not from their own country. Canada shows the same trend: the Diversity Index for Canada fell significantly from 42.25% at the country level to 14.08% at the continental level. Of the 71 faculty in Canada, 41 earned their PhD within the country, and 20 were educated in the United States. Conversely, the United States does not show a similar preference for hiring from its neighboring country. Though the U.S. Diversity Index decreased from 12.22% at country level to 8.89% at continental level, only 9 faculty received their PhD from Canadian institutions.

*Table 1. Diversity index across different levels. The dots following the country names represent the level of socio-economic development. Blue dots represent the Western countries, while red dots represent the Non-Western countries. The continental level diversity index measures the proportion of faculty whose PhD was earned outside their continent compared to the total number of faculty affiliated with institutions within that country. The country-level diversity index essentially measures the extent of domestic hiring versus bringing in talent from abroad. At the institutional level, the diversity index provides a measure of internal retention versus external recruitment. For example, In North America, 11.01% of current GIS faculty obtained their PhD from institutions outside of the continent. Within Canada, 14.08% of GIS faculty earned their PhD from institutions outside of North America, while 42.25% obtained their PhD*

*from institutions outside of Canada itself. At the institutional level within Canada, 88.73% of GIS faculty are employed at institutions different from where they received their PhD.*

| Continent | Country or Region | Continental Level | Country Level | Institutional level | Current GIS Faculty |
|---|---|---|---|---|---|
| North America 11.01% | Canada | 🔵 14.08% | 42.25% | 88.73% | 71 |
| | United States | 🔵 8.89% | 12.22% | 93.33% | 270 |
| Asia 24.37% | Singapore | 🔴 100.00% | 100.00% | 100.00% | 6 |
| | Thailand | 🔴 0.00% | 100.00% | 100.00% | 2 |
| | South Korea | 🔴 73.33% | 73.33% | 100.00% | 15 |
| | Malaysia | 🔴 66.67% | 66.67% | 66.67% | 3 |
| | Israel | 🔴 50.00% | 50.00% | 66.67% | 12 |
| | China | 🔴 20.16% | 20.55% | 73.52% | 253 |
| | Japan | 🔴 4.35% | 8.70% | 39.13% | 23 |
| Europe 11.15% | Ireland | 🔵 0.00% | 80.00% | 100.00% | 5 |
| | Norway | 🔵 22.22% | 55.56% | 66.67% | 9 |
| | Netherlands | 🔵 4.35% | 47.83% | 86.96% | 23 |
| | Switzerland | 🔵 13.04% | 43.48% | 47.83% | 23 |
| | Belgium | 🔵 20.00% | 40.00% | 40.00% | 5 |
| | Denmark | 🔵 0.00% | 36.36% | 36.36% | 11 |
| | Sweden | 🔵 11.11% | 33.33% | 50.00% | 18 |
| | Poland | 🔵 0.00% | 33.33% | 66.67% | 3 |
| | UK | 🔵 20.25% | 29.11% | 77.22% | 79 |

| | | | | | | |
|---|---|---|---|---|---|---|
| | Germany | 🔵 | 0.00% | 20.83% | 95.83% | 24 |
| | Austria | 🔵 | 0.00% | 12.50% | 31.25% | 16 |
| | Finland | 🔵 | 0.00% | 10.00% | 70.00% | 10 |
| | Czech Republic | 🔵 | 20.00% | 29.09% | 63.64% | 11 |
| | Italy | 🔵 | 0.00% | 0.00% | 50.00% | 4 |
| | France | 🔵 | 0.00% | 0.00% | 90.91% | 11 |
| | Estonia | 🔵 | 0.00% | 0.00% | 0.00% | 1 |
| Oceania 64.00% | New Zealand | 🔵 | 90.91% | 90.91% | 90.91% | 11 |
| | Australia | 🔵 | 38.46% | 46.15% | 53.85% | 13 |
| | Western | 🔵 | 13.17% | 32.15% | 65.01% | 618 |
| | Non-Western | 🔴 | 44.93% | 59.89% | 78.00% | 314 |

The analysis of diversity indices reveals significant disparities in the geographic representation of GIS faculty between Western and Non-Western countries. Our findings indicate that GIS scholarship is highly centralized in Western countries, leaving many Non-Western countries underrepresented. This trend aligns with the global distribution of geoscientists, as exemplified by the fact that countries such as the United States, the United Kingdom, and Germany accounted for over 60% of international collaborations in leading journals like Nature Geoscience between 2008 and 2017 ("Globalized Geoscience," 2018). Moreover, Western countries exhibit lower diversity indices across all levels, suggesting that faculty in these regions are more likely to have earned their PhDs within their own continent and that institutions tend to favor domestic hiring. In contrast, Non-Western countries demonstrate higher diversity indices, indicating a greater propensity to recruit faculty with international educational backgrounds.

*4.3  Language Congruency Analysis*

Language barriers are widely recognized as a factor influencing faculty placement. To examine this effect, we conducted a language congruency analysis comparing the primary language of instruction at each faculty member's PhD-granting institution with both the language(s) used at their current institution and the official language(s) of the region in which it is located. Our study did not incorporate each faculty member's first language due to the difficulty of tracking such information from public sources.

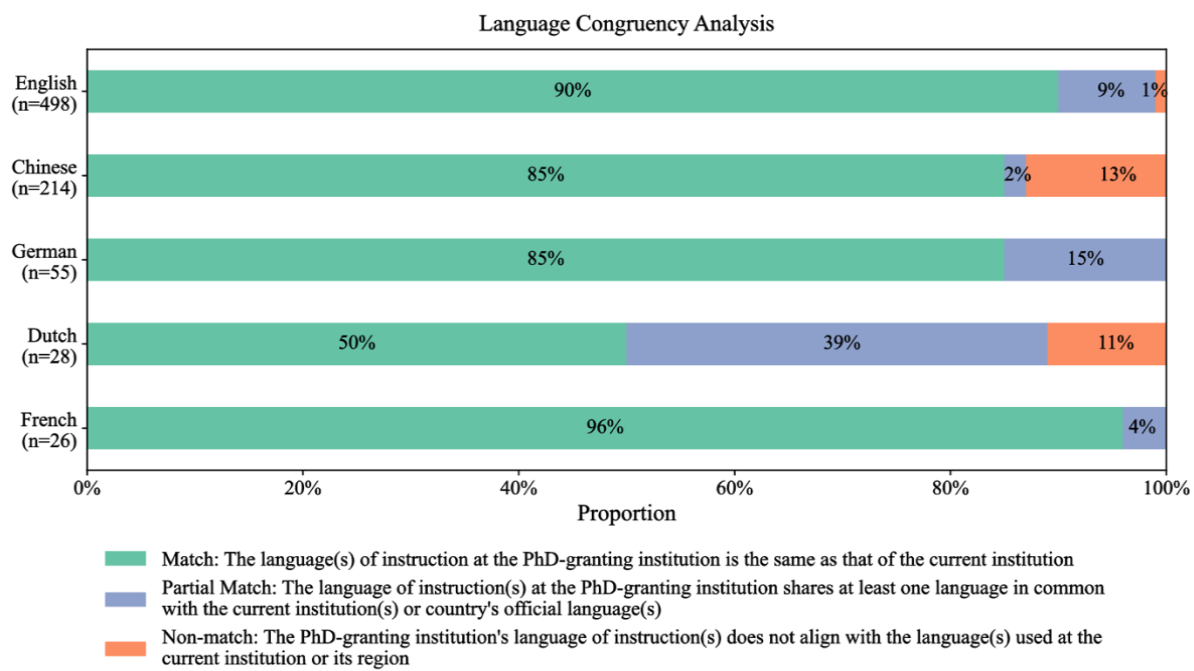

*Figure* 9. *Language Congruency Analysis.*

We focused on the top five languages with the highest institutional counts (*Figure* 9). Each university was annotated with its language(s) of instruction, primary language, and the regional official language. A placement was labeled as a "Match" if the languages were identical, "Partial Match" if there was some overlap between the language(s) of instruction or with the regional official language, and "Non-match" if no common language was found. Overall, each language group shows a "Match" rate of at least 50 percent, and when "Partial Match" is included, the combined figure exceeds 85 percent for every group. This suggests that, in the majority of cases, the PhD-granting institution and the hiring institution (or its region) share at least one common language, highlighting the significant role that linguistic compatibility plays in faculty placements. Chinese institutions exhibit the highest non-match rate (15%), many of these cases represent Chinese faculty who graduated from Western universities (e.g., the US and Canada) and returned to China. This outcome also reflects language congruency in a broader sense.

It is worth noting that, especially in many European countries where English is not the primary language, institutions often offer graduate-level courses in English or explicitly maintain an English-language research environment. This practice not only helps lower language barriers for scholars whose native language differs from the local medium of instruction, but also underscores the global importance of English in GIScience academic environment.

## 5. GIS Research Themes

After examining faculty hiring networks, we shift our focus to the research themes pursued by these faculty. Faculty placements are influenced not only by qualifications but also by the need to align institutional research goals with emerging trends. The focus areas of newly hired faculty help determine which sub-disciplines develop and how research priorities are set, thereby impacting future academic directions in their institutions. By analyzing research themes, we aim to reveal how hiring patterns contribute to the growth of specific areas within the discipline and reflect the underlying priorities of academic institutions. This analysis bridges the study of faculty placements with the broader narrative of knowledge transmission in GIScience.

### *5.1 Methods*

Identifying latent topics requires an understanding of the listed research interests of each scholar in the dataset and categorizing them into several classes. To achieve this goal, a four-step approach is taken, which involves (1) sentence embedding, (2) spatial clustering, (3) semantic analysis, and (4) data visualization.

Sentence embedding is a technique in natural language processing (NLP) that transforms a sentence into a fixed-length vector of numbers. These vectors (embeddings) capture the semantic meaning of the sentence or a list of words, which is the precondition of the downstream tasks such as sentence similarity analysis and sentence clustering (Bodrunova et al., 2020; Khan et al., 2020). To take advantage of the progress in NLP, we leveraged transformer-based large language models implemented by Langchain[4] to accomplish this task such that the embedding can capture the complex meanings of the research interests (Reimers & Gurevych, 2019).

Spatial clustering method is then applied to separate the high-dimension sentence embedding into several spatial clusters. K-means clustering, DBSCAN, Hierarchical clustering, and Spectral clustering are commonly used clustering methods to separate geospatial data. Spectral clustering (Von Luxburg, 2007) uses eigenvalues of the similarity matrix of the data to perform dimension reduction before clustering and is capable of capturing intricate relationships between points, which is good for non-convex clusters. It is suitable to the task of clustering sentence embedding and leveraged in this study.

The t-distributed Stochastic Neighbor Embedding (t-SNE) is a dimension reduction method frequently used in machine learning. It is particularly effective at preserving the local structure of data and revealing patterns in datasets with complex relationships (Van der Maaten & Hinton, 2008). In this study, t-SNE is used to reduce the high-dimension sentence embedding to two dimensions to visualize and evaluate the clustering results.

To better understand the semantic of each research cluster identified by the aforementioned methods, word cloud visualization and Latent Dirichlet Allocation (LDA) are utilized. Word cloud visualization is widely used in information extraction and keyword metadata depicting of social media posts (e.g., ref A, ref B, etc.) and was adopted in this study. It draws a collection of words in various sizes based on word frequency (Yum, 2020; Zhou et al., 2024). LDA is a Bayesian-based probabilistic unsupervised topic modeling algorithm that detects hidden topics regarded as a multinomial distribution over words (Blei, 2003). In other words, the detected topics in which high-frequency words are more likely to be present are the ones that occur more often in the collection of sentences.

---

[4] https://www.langchain.com/

## 5.2 Clusters Based on Placements

### 5.2.1 Clustering Results

After applying Transformer based large language model to the research interests of each faculty, the research interests are embedded to a 734-dimention vector. Then, the spectral clustering method is applied to group the vectors. The number of clusters, K, is the hyperparameter that needs to be tuned during the calculation process. Based on the GIST body of knowledge (*GIS&T Body of Knowledge Visualization and Search*, 2024), there are 10 knowledge areas, therefore we set the maximum value of K=10. We perform the clustering calculation iteratively with the value of K starting from 3 to 10. Since no ground truth data is available for the clustering task, visual inspection is applied to evaluate the impact of different K values on the clustering performance. Scatter plots of the clustering results are drawn after reducing the embedding to 2-dimension using t-SNE. We observed clear separation, compact and cohesive clusters, consistent shapes, and few overlapping or scattered clusters when K=7. The cluster result after dimension reduction is shown in Figure 10.

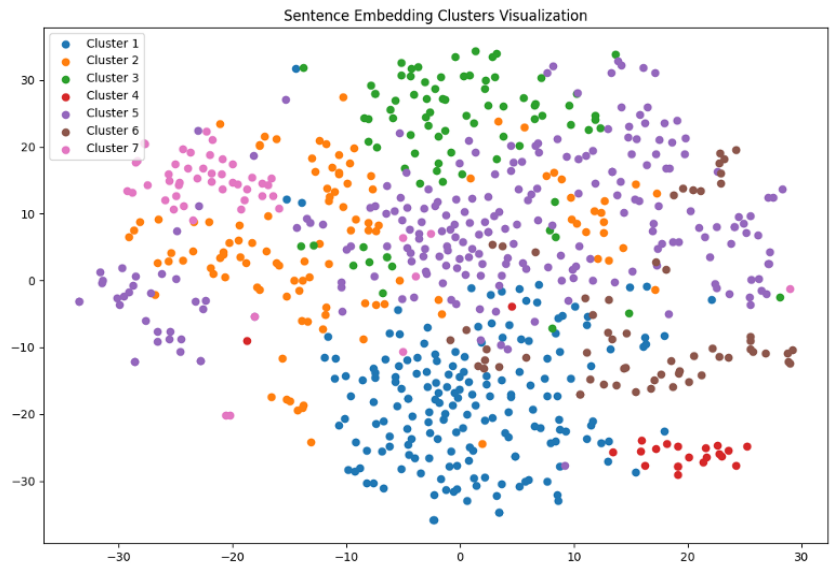

*Figure 10. Spectral clustering results of research interests after dimension reduction by t-SNE. Cluster 1: Environmental and ecological analysis, remote sensing; Cluster 2: Urban planning and social geography; Cluster 3: Cartography and visualization; Cluster 4: Geospatial data analysis and modeling; Cluster 5: Spatial analysis and urban systems; Cluster 6: 3D modeling; and Cluster 7: Health geography.*

### 5.2.2 Cluster Semantics

To have a more intuitive understanding of each research cluster, semantic analysis is performed using LDA and word cloud visualization. The result of the LDA model indicates a robust and reliable identification of distinct topics within the dataset. Each topic is characterized by a coherent set of high-probability words that collectively provide a clear thematic interpretation. The topics for each cluster are listed below:

Cluster1: Environmental and ecological analysis, remote sensing

Cluster2: Urban planning and social geography

Cluster3: Cartography and visualization

Cluster4: Geospatial data analysis and modeling

Cluster5: Spatial analysis and urban systems

Cluster6: 3D modeling

Cluster7: Health geography

Environmental and ecological analysis, urban planning and social geography, and geospatial data analysis and modelling are the largest three research clusters. The topics exhibit minimal overlap, suggesting well-defined boundaries between different thematic areas, which enhances the interpretability of the results. Additionally, the distribution of topics across documents is consistent with the expected thematic patterns, reinforcing the reliability of the LDA results. This outcome provides valuable insights for further analysis and interpretation.

The word cloud for each cluster is visualized in Figure 11 to further validate the result. The size of each word in the word cloud corresponds to its frequency or importance within the text. Larger words appear more frequently or hold greater significance, while smaller words are less common or less significant. The differences in color are intended to improve visual discriminability.

The result of the word clouds visualization provides an immediate visual summary of the most important terms and is coherent with the LDA results, which further demonstrates the reliability of the research cluster identified.

| Cluster 1 | Cluster 2 | Cluster 3 |
| --- | --- | --- |
| Environmental and ecological analysis, remote sensing | Urban planning and social geography | Cartography and visualization |

| Cluster 4 | Cluster 5 | Cluster 6 |
| --- | --- | --- |
| Geospatial data analysis and modeling | Spatial analysis and urban systems | 3D modeling |

Cluster 7
Health geography

*Figure* 11. *Word cloud visualization for each research cluster.*

## 5.3 Clusters Based on Semantic Word cloud

As an interdisciplinary field, GIScience has evolved from a focus on GIS applications to a flourishing scientific discipline, experiencing significant expansion in its sub-domains over the past three decades. To trace these developments, we analyzed the research interests of faculty using semantic word clouds. Out of 946 faculty members, 755 had listed their PhD graduation years on their profiles. We classified them into four groups according to their graduation periods to examine their research focuses, the time periods were chosen to reflect significant phases in the evolution of GIScience. *Figure* 12 presents the research interests of faculty who completed their PhDs within the following periods: 1990 and earlier (52), 1991-2000 (161), 2001-2010 (299), and 2011 onwards (243).

In our analysis, we excluded broad research themes like spatial analysis, spatial data science, and Geographic Information Systems, aiming to explore the evolution of sub-topics or cross-discipline topics.

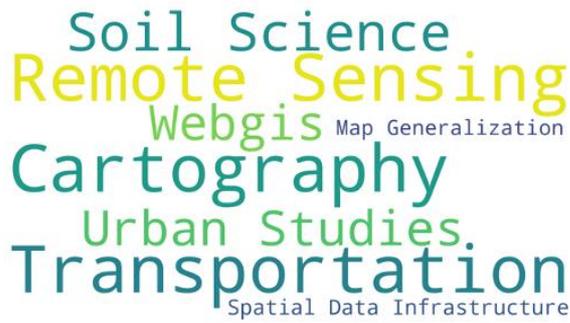
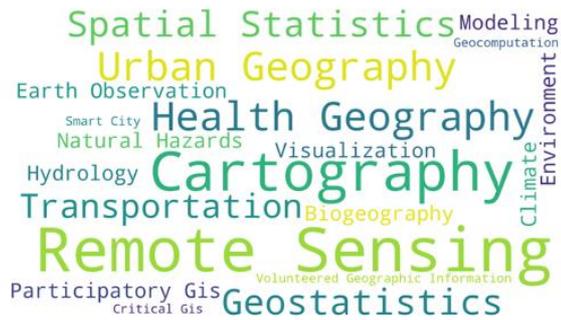

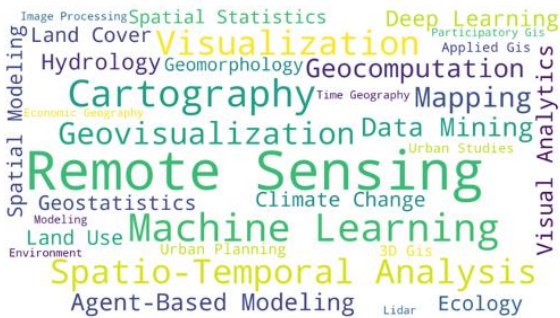
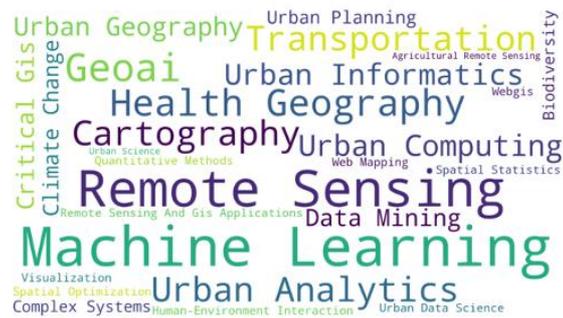

*Figure* 12. *Word cloud visualization of the research interests of faculty who completed their PhDs across four periods.*

Overall, our analysis highlights four notable trends in GIS-related research interests over time. First, beyond the consistently present themes such as remote sensing, cartography, and transportation, different focuses emerge across different groups. Faculty who graduated from 1991 to 2000 possessed research interests that emphasize natural studies, such as biogeography, hydrology, and climate, reflecting the integration of GIScience with environmental disciplines. Among the third group (2001-2010), cartography-related sub-topics were frequently mentioned, including mapping, geovisualization, visualization, and visual analytics. Since 2011, the focus has pivoted to more detailed urban issues, with topics such as urban computing, urban analytics, and urban informatics being particularly prominent. Second, there is a clear trajectory toward increasingly interdisciplinary research interests. Faculty with PhDs before 1990 concentrated on soil science, urban studies, and transportation. From 1991-2000, the scope expanded to include environmental science topics. After 2000, the diversity of interests grew further to include research on geoscience and human geography. Third, computational technology has emerged as a significant focus since 2001. Machine learning, data mining, and geocomputation were listed between 2001 and 2010, and were further branched into specialized themes in Geography such as GeoAI, and urban computing from 2011 to present. This trend parallels the rise of AI since the early 2000s. Fourth, the frequent mention of participatory GIS, Volunteered Geographic Information (VGI), and critical GIS indicates a growing interest in placing humans at the center of GIS inquiries.

## 6. Discussion

This study examines global GIS faculty placements and research interests by analyzing faculty hiring patterns at continental, country, and institutional levels. We list several takeaways that may offer insights for future students and the broader GIS community.

### *6.1 High Internal Recruitment at Continent Level and Country Level*

Analyzing faculty hiring at the continental and country levels reveals a tendency toward internal recruitment. In other words, academic institutions prefer to hire faculty who obtained their PhD within the same country or continent, reflecting a strong inclination to hire local talent. Several factors may explain this pattern:

**Institutions' Legacy.** Countries with long-standing academic traditions often develop strong educational infrastructures that can favor the development of faculty from within their own ranks, potentially reducing the reliance on academics from outside. For instance, well-established universities tend to build extensive networks of alumni who continue their careers within the same national or regional context. This can create a self-sustaining cycle where local graduates are more likely to be hired as faculty, reinforcing the internal recruitment trend. Additionally, these institutions often have established pipelines for nurturing academic talents from undergraduate through to postgraduate levels, which further entrenches internal recruitment.

**Geopolitical Dynamics.** The state of a country's diplomatic relationships and trade agreements can shape the movement of academic talent. Countries that maintain open and strong international connections tend to experience more academic exchanges. In contrast, countries with restrictive immigration policies or strained international relations may see a reduced influx of international faculty. For example, countries within the European Union (EU) benefit from policies that facilitate the free movement of people. Based on our diversity index analysis, the index for nine European countries at the continental level is 0%, indicating that these countries predominantly hire faculty from within Europe. This finding highlights the effectiveness of intra-European policies, such as free movement agreements, which not only facilitate academic exchanges within Europe but also tend to retain European scholars on the continent, thereby limiting cross-continental mobility. Due to free movement, geographic proximity, and strong intra-regional networks (e.g., among DACH or Benelux countries), the EU often functions as a single integrated market. An international move within the EU—such as between Germany and Austria—is less significant than moves between regions with greater barriers, such as between Germany and the United Kingdom or between an EU country and a non-EU country like New Zealand.

**Geographical Proximity.** Canada's diversity index drops from 42.25% at the country level to 14.08% at the continental level, indicating heavy reliance on the U.S. for academic talents. Similarly, in Europe, countries like Ireland, Norway, the Netherlands, and Switzerland experience substantial drops in their diversity indices—each exceeding 20%—when shifting from the country level to the continental level, underscoring the importance of geographically neighboring countries in academic hiring.

**Financial Considerations**. The expenses associated with campus visits for potential hires, which cover accommodation and travel, can be substantially higher for international candidates. This might influence the preference for candidates within the country or from geographically neighboring countries to manage costs. Institutions operating on limited budgets might prioritize local candidates to minimize recruitment expenses. Moreover, funding bodies and grants often have stipulations that favor hiring local talent, further incentivizing institutions to look inward rather than abroad. This financial prudence can lead to a concentration of hiring within specific regions, impacting the diversity of academic staff.

**Cultural Barriers.** Language differences and varying academic traditions can impede collaboration among researchers across regions. Our language congruency analysis indicates that scholars are more likely to remain at universities where there is at least one common language shared between their PhD institution or the current region. From the scholars' perspective, cultural barriers often lead them to actively choose familiar linguistic environments. From the institutions' perspective, such preferences can foster more localized academic networks in regions characterized by cultural and linguistic homogeneity, potentially restricting the broader circulation of knowledge.

**Personal factors**, including professional networks and institutional ties, could also influence where and how scholars pursue their careers. Many researchers rely on their established networks when making career decisions, such as connections with mentors, colleagues, and alumni. These networks provide insider information and opportunities for collaboration, often making it more attractive for scholars to remain within familiar institutions or regions. As a result, internal recruitment is reinforced. This localized hiring can, in turn, influence the distribution of research interests, when a particular region or institution is dominated by specific academic traditions or research themes, new hires are likely to align with those established paradigms.

*6.2 High External Recruitment at Institutional Level*

Previously, the absence of unified data limited the study of global GIS faculty mobility. The GISphere database now facilitates these investigations, marking a significant contribution to the field. Our institutional analysis showed that institutions generally prefer hiring faculty from outside their own system to foster intellectual diversity and avoid academic inbreeding. These non-self-hire practices are particularly notable in the United States. In our data, non-self-hiring accounts for approximately 94.59% of all U.S. GIS professors, the highest proportion among the top four countries with the most current GIS faculty. This observation aligns with prior studies on non-self-hire trends across countries (Wapman et al. 2022).

Hiring externally may bring in new perspectives and expertise, which can invigorate academic programs and stimulate innovative research. For example, faculty from diverse academic backgrounds can enhance the overall quality of research and teaching within an institution (Wiggins & Sawyer, 2012). Additionally, external hires can mitigate the risks of groupthink (Klein & Stern, 2009), which can occur when an institution relies heavily on internally developed faculty. Groupthink can stifle creativity and limit the diversity of thought necessary for addressing complex scientific research problems. By incorporating faculty from various academic and cultural backgrounds, institutions can foster a more dynamic academic environment.

*6.3 Global GIS Program and Overall Trend*

Certain institutions with influential GIS programs have been particularly successful at placing their graduates in faculty positions at other universities, such as Wuhan university, Chinese Academy of Science, and University of California Santa Barbara. It is encouraging to observe that, compared with fields such as Computer Science and Urban Planning — where a larger share of placements is concentrated among a few institutions — GIScience exhibits a more equitable distribution of faculty positions. The global leading five universities account for only 15.43% of such placements, implying a balanced distribution in GIScience. We advocate for sustaining the equality in academic hiring within the field. It is also important to recognize that external factors, such as the quality, quantity, and size of academic programs, also influence faculty placements. Universities that offer higher-quality programs and maintain robust research environments are likely to produce more competitive graduates, thereby naturally leading to higher placement rates. For example, according to the QS World University Rankings for Geography 2024 (2025), institutions in the United Kingdom and the United States each account for 19% of the top 100 universities, following by China mainland (7%). This distribution aligns with our analysis, as a higher proportion of GIS faculty originate from these countries. Furthermore, the availability of PhD programs is not uniform across countries—some countries lack sufficient PhD programs or have few of them, which further influences the observed placement patterns. Future studies could investigate changes over longer periods, such as 10, 20, and even 30 years, to determine whether faculty hiring in GIScience exhibits patterns of academic inequity. This longitudinal approach would allow for a more comprehensive understanding of hiring trends and whether they lead to an imbalance in academic opportunities over time. Moreover, the observed clusters and cross-border placements highlight the importance of interdisciplinary work in the field. The GIS community benefits from a diverse and inclusive academic environment that encourages collaboration and innovation. By examining the factors that contribute to successful faculty placements, such as research output, industry partnerships, and international collaborations, we can better understand how to create equitable opportunities for all graduates. A detailed analysis of research clusters based on faculty research interests can reveal whether certain sub-disciplines within GIScience are more prone to concentration in a few institutions. Identifying these patterns could facilitate a more balanced distribution of academic opportunities. For instance, promoting mentorship programs and collaborative research projects can enhance academic diversity and inclusivity in GIScience.

The dominance of Western countries, particularly the United States, in the GIS academic hiring market highlights a significant imbalance in the global distribution of GIS expertise, with many non-Western countries remaining notably underrepresented. The U.S. accounts for 28.54% of current global GIS faculty, with 40.49% of GIS faculty earning their PhD degrees in the U.S. Furthermore, U.S. institutions represent 40.63% of the top 25% of educational institutions in terms of GIS faculty placement. This imbalance highlights the need for intentional efforts to support GIScience PhD candidates from developing countries and to strengthen GIS programs within these regions, fostering a more inclusive and globally representative academic community.

*6.4 Shifting Focus and Emerging Trends in GIS Research Clusters*

The clustering results highlight the diversity of research interests among GIS faculty, with seven well-defined clusters identified: Environmental and ecological analysis, Urban planning and social geography, Cartography and visualization, Geospatial data analysis and modeling, Spatial analysis and urban systems, 3D modeling, and Health geography. These clusters exhibit minimal overlap, indicating clear boundaries between thematic areas and suggesting that GIScience is becoming increasingly specialized. The largest clusters, namely Environmental and Ecological Analysis, Urban and Social Geography, and Geospatial Data Analysis and Modeling, underscore the strong emphasis on environmental and urban issues, reflecting the field's response to pressing global challenges.

The temporal analysis of research interests through semantic word clouds reveals significant shifts in the focus of GIS research across different academic generations. Earlier cohorts, particularly those graduating before 2000, were primarily focused on natural and environmental sciences, reflecting the early integration of GIS technology with ecological and hydrological studies. As the field matured, there was a shift towards urban-related topics and the adoption of advanced computational techniques, such as machine learning, geocomputation, and GeoAI, particularly among those who graduated after 2000. This evolution demonstrates the increasing interdisciplinarity of GIS research, with a growing emphasis on urban informatics and the application of GIS in understanding complex human-environment interactions. Additionally, the rise of participatory GIS and critical GIS themes indicates a trend towards incorporating social dimensions and human-centric approaches within GIScience.

In conclusion, addressing the concentration of faculty placements in influential GIS programs is crucial for ensuring a diverse and equitable academic environment. By investigating long-term trends and fostering interdisciplinary and international collaborations, the GIS community can promote a more inclusive future for the development of GIScience.

### *6.5 Limitations and Future Work*

We acknowledge several limitations in our study. The first issue relates to the scope of our database. Despite our efforts to comprehensively collect data on GIS faculty worldwide, it is likely that we did not capture all relevant individuals. This oversight may be attribute to several factors, including missing institutions or GIS-related programs, and the absence of certain faculty on departmental website. There may be inherent biases in the dataset due to the geographic distribution of the volunteers who contributed to its creation. Since many of the volunteers are more familiar with institutions in the North America and Europe and other well-represented regions, the database might have more complete information on faculty from these areas, potentially underrepresenting GIS scholars from institutions in non-Western countries, especially for institutions in Africa and South America. In our attempts to collect information from non-Western countries, we also encountered challenges such as language barriers and the lack of publicly available data. In future work, we plan to address this limitation by leveraging participatory GIS approaches to enhance data collection from underrepresented regions. Furthermore, the reliance on publicly available information may skew the dataset toward institutions that are more transparent and active in maintaining accessible online profiles for their faculty. In addition, while our analysis centers on doctoral training due to data consistency, it is important to recognize that postdoctoral experiences and a myriad of other personal and professional factors frequently play a crucial role in determining faculty trajectories. Despite these limitations, we believe that the data collected captures key trends within GIScience faculty hiring. Our study provides detailed analysis of global placement patterns and highlights the role of academic networks in shaping the dissemination of GIScience knowledge.

Second, despite our efforts to refine our inclusion criteria, our operational definition of GIS faculty, which is based on research interests and publication records, may inadvertently include individuals who primarily identify as experts in other fields, even though they have conducted research related to GIS. Conversely, the inherent ambiguity and conflicting terminology within GIScience may have resulted in the inadvertent exclusion of some relevant faculty. Moreover, our focus on tenure-track faculty may introduce biases by overlooking the contributions of non-tenure-track professionals, who also play a significant role in the field. It is also important to note that the concept of "tenure track" varies across countries and institutions, with some not offering tenure-track positions at all or employing different criteria and guidelines for tenure. Nevertheless, we believe that our results provide a relatively comprehensive and objective reflection of faculty placement in GIS.

Third, our network analysis of faculty placement focuses solely on comparing faculty members' PhD institutions with their current affiliations, and thus does not thoroughly examine their full academic trajectories—such as their bachelor's and master's degrees or past appointments. Future studies could aim to construct a more detailed picture of the academic trajectories of GIS faculty. Additionally, our analysis does not account for other factors that may influence faculty placements, such as individual qualifications, demographics, or the size of PhD programs. Larger programs are likely to produce more PhD graduates, which may affect their representation in faculty placements. However, PhD program size is inherently variable, and departments rarely disclose specific figures. Consequently, using the number of faculty originated as a proxy measure is a common practice in faculty placement studies (Clauset et al., 2015; C. A. Lee, 2022). Further research could investigate these factors to gain deeper insights into the decision-making processes behind GIS faculty appointments. A further limitation of our study is that we lack data on the applicant pool for these positions; if international candidates apply in lower volumes than domestic candidates, this could partly explain the higher percentage of domestic hires, independent of internal gatekeeping or bias. Moreover, while we identified influential GIS institutions based on faculty placement in PhD programs, academia represents only one career path for students. Many students may opt for careers in the industry and choose to enroll in institutions that offer professional GIS programs or master's degrees instead of pursuing PhD degrees. Consequently, it is possible that some institutions that supply strong candidates to the GIS industry are overlooked in this study. In the future, we aim to explore influential programs within the industry to see if they reflect the trends observed in academia.

## Disclosure statement

No potential competing interest was reported by the author(s).

## Acknowledgement

The authors would like to thank all volunteers and contributors to the GISphere project and its community for their valuable efforts and support. We also extend our sincere gratitude to the International Association of Chinese Professionals in Geographic Information Sciences (CPGIS) and the University Consortium for Geographic Information Science (UCGIS) for their generous support and encouragement throughout this work.

**Funding**

This work is supported by the Cartography and Geographic Information Society (CaGIS) Rising Award and GISphere Corporation.

**Data and Codes Availability Statement**

The data and codes that support the findings of this study are available with the identifier(s) at the private link (https://figshare.com/s/2cad8a9ec05b7cd184f5).